\documentclass[12pt]{article}

\begin{document}

%\draft
\title{Isotopic and spin selectivity of H$_2$ adsorbed in bundles of carbon nanotubes}
\maketitle
\author{R.A. Trasca, M.K. Kostov and  M.W. Cole \\{\it Department of Physics, The Pennsylvania State University, University Park, PA 16802}}

\begin{abstract}
Due to its large surface area and strongly attractive potential, a bundle of carbon nanotubes is an ideal substrate material for gas storage. In addition, adsorption in nanotubes can be exploited
in order to separate the components of a mixture. In this paper, we investigate the preferential
adsorption of $D_2$ versus $H_2$ (isotope selectivity) and of ortho versus para (spin
selectivity) molecules confined in the one-dimensional grooves and interstitial channels
of carbon nanotube bundles. We perform selectivity calculations in the low
coverage regime, neglecting interactions between adsorbate molecules. We find substantial spin selectivity for a
range of temperatures up to 100 K, and even greater isotope selectivity for an extended range of
temperatures, up to 300 K. This isotope selectivity is consistent with recent experimental
data, which exhibit a large difference between the isosteric heats of $D_2$ and $H_2$
adsorbed in these bundles.
\end{abstract}

%\maketitle

\section {Introduction}

    The adsorption of gases within single-walled, carbon nanotubes (SWNT)
has recently attracted broad attention among physicists, chemists, materials scientists
and engineers \cite{heb,che,sma,jo1,jo2,jo3,mer,ino,gao,yin,kuz,wil,dre,sta,col,sta2,cheng}.
Experiments have shown
that one can create ordered arrays of nanotubes \cite{the,sch,sin,gou,luc}, which
form a close-packed bundle (or rope) of SWNT's. Ideal bundles
of nanotubes consist of very long strands of nearly parallel tubes held
together at equilibrium separation by intermolecular forces.  The
small diameter and large aspect ratio of the nanotubes
make them interesting systems for gas adsorption. In
such a bundle, physisorption may occur at four distinct sites: i)
inside the tubes (endohedral adsorption); ii) in the interstitial
channels (IC) between three contiguous nanotubes (exohedral adsorption);
iii) in the grooves between adjacent nanotubes on the external surface
of the bundle, and iv) elsewhere on the outside surface of the bundle. In this paper, we assess
the ability of nanotube bundles to preferentially
adsorb specific isotope and spin species.

Small atoms and molecules, in contrast to larger molecules, have been predicted to fit well in the ICs of
a nanotube bundle \cite{the,kos}. For such species, the IC's environment provides a large
number of neighboring C atoms at nearly optimal distance from the adatom, so the
interstitial binding energy is larger than is found in other known
environments \cite{sta2}. The grooves also present strongly attractive environments for
both small and large molecules. The binding energies in these sites have been determined experimentally
to be between 50 \% and 100 \% greater than the values on graphite \cite{mig,heb,ye,wils}. 

Molecular hydrogen inside SWNTs is a particularly appealing system to study,
since its adsorption in ICs and grooves is relevant to both gas storage and
isotope separation \cite{jo5,gor,han,bas}. ``Selectivity'' is the term
used to describe the separation, or selective adsorption, of one species relative 
to the other species of a mixture. In a mixture of two
components, the selectivity of component 1 relative to that of 2 is:
\begin{equation}
S=(x_1/x_2)/(y_1/y_2)
\end{equation}
with $x_i (y_i)$ the pore (bulk) molar fractions. Since two isotopes have similar sizes,
shapes and interaction potentials, the separation of isotopic mixtures is a
difficult and energy-expensive process, requiring special experimental
techniques, such as cryogenic  distillation, diffusion separation,
laser isotope separation or microwave molecular separation
\cite{enc,green,ave}. However, most of these processes have low selectivity for
separating $H_2$ isotopes.
Recently, a
novel separation technique called quantum sieving was predicted to be particularly
efficient for nanotubes \cite{jo4,jo5,bee,hath}. Quantum sieving separates lighter molecules from heavier
ones by selectively adsorbing heavier molecules. Such  selective
adsorption can be explained by the higher zero-point energy of
the light species, which makes their adsorption relatively
unfavorable. Quantum sieving can be implemented when the adsorbate is
effectively confined to a one-dimensional (1D) channel or a 2D surface or a small
cavity (0D). 
The selectivities observed for adsorption of
$H_2$ isotopes on common substrates such as graphite, zeolites, and
alumina are low - typically in the range of 1.1 to 3 \cite{bas,bas1,steph}.
However, differences in zero-point energies of the adsorbed
species are expected to be particularly large when molecules are confined in
very narrow pores. Wang {\it et al.} have shown that nanopores with
diameters wider than 7 $\mathring A$ exhibit weak selectivity
while smaller nanopores were predicted to exhibit large selectivities \cite{jo4}. Carbon
nanotubes typically have diameters larger than 7 $\mathring A$, but the interstitial
channels are smaller, so that they have a pore size
and solid-fluid
potential which can effectively sieve mixtures of $H_2$ and $D_2$. Indeed, Path Integral Monte Carlo
calculations at variable pressures \cite{jo5} yielded large isotope selectivity
in the IC. Our calculations are performed in the limit of low coverage (virtually zero pressure), 
using what we believe to be an improved gas-solid interaction potential. 

 The solid-fluid potential
is usually modelled as a pair-wise sum of Lennard-Jones interactions between adatoms and
carbon atoms \cite{jo5,gor}. The potential used in our study is the one recently developed by 
Kostov {\it et al.} \cite{kos2}. In addition to terms present in previous calculations, 
Kostov's potential includes also the
interactions of the $H_2$  quadrupole moment with the local electrostatic field of C
nanotubes (found from ab initio
calculations \cite{drag}) and interactions of the $H_2$ static multipole moments with the image
charges induced on the surrounding nanotubes. The model yields a strongly confining potential for
$H_2$ and $D_2$, of depth about 1800 K ($\sim$ 160 meV), which alters the rotational spectrum and induces a large
difference between isotopic zero point energies. As we will show, the calculated difference between $H_2$ and
$D_2$ binding energies is consistent with recent isosteric heat data of Wilson {\it et al.}
\cite{wils}. This large difference, when exponentiated in an Arrhenius expression at
low temperature, yields huge isotope selectivities
in the IC and groove, far exceeding that on common sorbents (e.g. graphite and
zeolites). 

   The rotational hindrance of $H_2$
   molecules adsorbed in carbon nanotube bundles is a controversial issue. There exist
   experiments showing no alteration of the rotational spectrum \cite{nist} and other
   experiments (with different samples) which
   manifest large changes compared to the free rotation energy spectrum \cite{nare}. The
   calculations of Kostov {\it et al.} indicate
   that interstitially adsorbed $H_2$ has a significantly
   hindered rotational motion \cite{kos2,nare}. The existence of a large
   rotational barrier leads to large splittings and shifts (relative to an orientationally
   invariant potential) of the $H_2$ and $D_2$
   rotational energy levels (see Fig.1). In this work we show that a consequence of
   this hindered rotational motion is a significant spin
   selectivity in the ICs and grooves.

   %In general, the quantization of the
  % rotational degrees of freedom of confined molecules can lead to isotopic effects
   %not present in the bulk phase. 
   Recently, Hathorn {\it et al.} showed that
   quantization of the restricted rotational motion of $H_2$ and $D_2$ confined in
   SWNTs contributes significantly to quantum sieving \cite{hath}. Their study, however, is
   substantially different from the present work, since they consider $H_2$
   adsorbed within a single nanotube and approximate the rotational potential as
   that of the molecule on the axis of the nanotube. In addition, they employed a
   modified Lennard-Jones potential with $\epsilon$ and $\sigma$ parameters far
   different from the potential parameters assumed in this work.
   They find significantly hindered rotational spectrum for the adsorbed
   molecule for small nanotubes. Their results are qualitatively consistent with
   our results although we are considering quite different sites.

The outline of this paper is as follows. In the next section, we
calculate the ortho-para selectivity as a function of temperature (T),
taking into account the hindered rotational motion. In
Section III, we derive the specific heat due to the rotational motion of the
adsorbed molecules. Section IV presents results for the
isotope selectivity of mixtures of 
$H_2$ and $D_2$. Section V presents calculations of the isosteric heats of 
$H_2$ and $D_2$ and compares these with the experimental
results of Wilson {\it et al.} \cite{wils}. Section VI summarizes and discusses
this work.

\section {Ortho-para selectivity}

In order to compute the selectivity of component 1 (e.g. ortho) relative to component 2 (e.g. para)
at low coverage, we use the equilibrium condition relating the chemical potential of  each component of the 3D vapor(outside the nanotube bundles) to that of  the 1D gas adsorbed in the IC:
\begin{equation}
\mu^{vapor}_i=\mu^{IC}_i
\end{equation}
where i stands for each component, ortho and para.
We employ statistical mechanics to find the chemical potential $\mu=(\partial F/\partial N)_{T,V}$,
where $F=-k_B T \ln Q$ is the Helmholtz free energy, $Q$ is the canonical partition function and $k_B$ is
Boltzmann's constant. In the low
density limit, the partition function involves energy contributions from (hindered) rotation, vibration and from translation  of the center of mass (CM) in the presence of a transverse confining potential.

 In the IC or groove,
the molecules are assumed to move freely  along the axial direction, so the
fluid is described
by a constant(mean) 1D density, $\rho$. The effect of corrugation on the motion is sensitive to
assumptions about registry of adjacent tubes \cite{milton,bon} and is neglected here. 
The CM transverse oscillation is described by the one-particle partition
function, $q^{cm}$. The internal degrees of freedom consist of the intra-molecular
vibrations and hindered rotations, described by $q^{vib}$ and $q^{rot}$, respectively.
Therefore:
\begin{equation}
F_i^{IC}= -N k_B T \ln[\frac{e}{\rho_i \lambda} q_i^{cm} q_i^{vib} (g_i q^{rot}_{i}) \exp(- C_i/(k_B T))]
\end{equation}
\begin{equation}
\mu_i^{IC}=C_i- k_B T \ln(\frac{e}{\rho_i \lambda})-k_B T \ln {q_i^{cm}}-k_B T \ln {q_i^{vib}}-k_B T \ln {(g_i q_i^{rot})}
\end{equation}
where i stands for para and ortho,
$\lambda=\sqrt{2\pi \hbar^2/(m k_B T)}$, $g_i$ is the nuclear spin degeneracy and $C_i$
is the interaction energy experienced by molecules due to the confining environment.

The partition function of the coexisting gas outside the nanotube bundle
has contributions  from the 3D translation motion and
from internal degrees of freedom (vibration and free rotation of the molecules). The free
energy and the chemical potential are then:
\begin{equation}
F_i^{vapor}=-N k_B T \ln[\frac{e}{n_i \lambda^3} q_i^{vib} (g_i q_{i}^{free})]
\end{equation}
\begin{equation}
\mu_i^{vapor}=-k_B T \ln{\frac{e}{n_i \lambda^3}}-k_B T \ln {q_i^{vib}}-k_B T \ln{(g_i q_i^{free})}
\end{equation}
where $n_i$ is the 3D density of ortho or para vapor outside the nanotube bundle.

Experimental data of Williams {\it et al} and
calculations of Kostov {\it et al} indicate
that the intra-molecular
vibrations are essentially the same in the IC as in free space \cite{kos,fitz}.
The CM oscillations depend
only on the molecular mass, so they give the same contribution to the
ortho and para partition functions. Finally, the only relevant contributions to the spin
selectivity come from
the even and odd rotational partition functions, denoted $q_+$ and $q_-$, respectively:
\begin{equation}
q_{+} = \sum_{j=even}\, \sum_{m=-j}^j g_{jm} \exp{(-\beta \epsilon_{jm})}
\end{equation}
where $\beta=1/(k_B T)$, $g_{jm}$ is the rotational degeneracy (not including the nuclear degeneracy), and $\epsilon_{jm}$ are the rotational
levels. For the odd partition function ($q_-$), the summation in $Eq.7$ is done only over the odd
rotational levels. In free space, both $H_2$ and $D_2$ can be modelled as free quantum rotors with 
$g_{jm}=j(j+1)$ and
$\epsilon_{jm}=j(j+1)B$. Here j is  the rotational quantum number and $B=\hbar^2/(2I)$, the rotational
constant of $H_2$ or $D_2$.  However, the different character of the
nuclear spin (fermionic for $H_2$ and bosonic for $D_2$) imposes different conditions on the
total wavefunction (antisymmetric for $H_2$ and symmetric for $D_2$), so that the nuclear 
degeneracies corresponding to even and odd rotational states are different: $g_+^{H_2}=1$,
$g_-^{H_2}=3$ and $g_+^{D_2}=6$, $g_-^{D_2}=3$.
By convention, the name ortho is given to
the species with larger statistical weight, so the even states are called para for $H_2$
and ortho for $D_2$. Thus, at high T, the ortho
and para molecules exist in the ratio of 3:1 in the case of $H_2$ and 6:3 in the case of
$D_2$.

The
spin selectivity in the low coverage (pressure) limit involves a ratio of densities,
which can be found by imposing the equilibrium condition ($Eq.2$): 
\begin{equation}
S=\frac{(\rho_-/\rho_+)_{IC}}{(n_-/n_+)_{vapor}}
\end{equation}
where $-$ and $+$ subscripts correspond to the odd and even rotational states,
respectively (ortho and para for $H_2$, para and ortho for $D_2$). 
 All similar factors mentioned previously cancel out in $S$ and,
in the end, the low coverage selectivity depends
on just the even and odd partition functions, $q_+$
and $q_-$:
\begin{equation}
S= \frac{(q_-^{IC}/q_+^{IC})}{(q_-^{vapor}/q_+^{vapor})}
\end{equation}

A qualitative understanding of the spin selectivity can be achieved by
examining the rotational  spectrum.
The ortho-para ratio in the IC, in the limit of high T, is the same as in the gas phase. But
since the energy spectrum changes in the IC, the overall partition functions differ,
with particularly large effects at low T.
Fig.1 contrasts the
rotational spectra of $H_2$ in the IC (as computed by Kostov) and in free space. In the IC, the reference energy is
the energy the molecule would have if it does not rotate ($C_{H_2}$=-1275 K)\cite{energy}. Notice that
the ground (para) state is shifted down from this value by $\sim$ 200 K and the first excited state (ortho
$j=1,m_j=0$) is also shifted down ($\sim$ 60 $K$), close to the ground state. The $D_2$ rotational
spectrum undergoes a similar alteration in the IC, but with a smaller shift of the ground (ortho)
state (about 130 K) with respect to a different reference energy ($C_{D_2}$=-1445 K),
and a slightly larger excitation energy to the ($j=1, m_j=0$) para state
(about 85 K)\cite{hath2}.
% The reference energies differ for $H_2$ and $D_2$ \cite{energy}. 
%Similar shifts and splittings of the rotational spectrum
%have been found previously by Hathorn {\it et al} inside nanotubes
%\cite{hath}.
The shift in the first
excited state (ortho for $H_2$, para for $D_2$) towards the ground state (para for $H_2$,
ortho for $D_2$) causes the ortho to para ratio of $H_2$ molecules, or para to ortho ratio
of $D_2$ molecules,
to increase substantially in the IC with respect to that in free space. In the groove site,
these ratios are still larger than 
in free space, but smaller, however, than in the IC, due to the lesser confinement between the two
nanotubes in the
groove site.
At low T, all but the lowest states yield negligible contributions to $q^{rot}$, so that
the selectivity depends only on the values of the
lowest excitation energies in the IC and free space.
\begin{equation}
S_o=\frac{\exp[-\beta (\epsilon_{10}-\epsilon_{00})]+2\exp[-\beta (\epsilon_{11}-\epsilon_{00})]}{3\exp[-2 \beta B]}
\end{equation}
where the free space separation is twice the rotational constant 
B=85 K for $H_2$.

The results for the ortho-para selectivity are shown in Fig.2. We find that the spin
selectivity is larger than 1 for an extended range of temperatures, up to 100 K in the IC
and $\sim$ 75 K in grooves. This
is consistent with the difference between the excitation energies in the IC (groove)
and in free space which enters $Eq.10$.
As discussed above, the alterations of the rotational
spectrum favor the $H_2$ ortho species,
whereas for $D_2$, the para molecules are preferentially
adsorbed. Since the $H_2$ molecules
have a larger zero point motion than $D_2$ molecules, they experience more rotational
hindrance and their spectrum is more altered in the corresponding energy
scale. Since the reference energy does not count for the spin selectivity, this
exhibits larger values for $H_2$ than for $D_2$.

\section{Rotational specific heat}

In order to compute the thermodynamic properties of the adsorbate, one works with 
the composite equilibrium partition function:

\begin{equation}
q^{H_2}_{rot}=S(2S+1)q_{+}+(S+1)(2S+1)q_{-}
\end{equation}
\begin{equation}
q^{D_2}_{rot}=(S+1)(2S+1)q_{+}+S(2S+1)q_{-}
\end{equation}
where $S=1/2$ for $H_2$ and $S=1$ for $D_2$. We will discuss the $H_2$ case 
and point out the differences in the case of $D_2$.
In many  circumstances, $H_2$ is not in thermal equilibrium as regards the 
relative magnitudes of para and ortho components, because the probability of flipping the
nuclear spin is very small (the lifetime in free space is $\sim$ one year).
Therefore, the transition probability of a molecule from one nuclear spin state to another is
negligible during a specific heat experiment.
Consequently, the sample may be viewed as a non-equilibrium mixture of two independent
species, which give additive contributions to the heat capacity. 
\begin{equation}
C_{non-eq}=f_p C_{p} +f_o C_{o}
\end {equation}
Here $f_p$ and $f_o$ are the para and ortho molar fractions. In an equilibrium
mixture these depend on temperature, whereas in a non-equilibrium mixture their relative
concentrations are determined by the initial conditions. In our calculations, we let the para and ortho 
molar fractions have  their high T values: $f_p=1/4$ and $f_o=3/4$.
The para and ortho specific heats (in the low
density limit) are:
\begin{equation}
C_{p,o}=N_{p,o} k_B\frac{d}{dT}(T^2 \frac{d}{dT} \ln{q_{+,-}})
\end{equation}

At low T, only the lowest levels contribute to the specific heat. Let us consider only the first
two levels: $\epsilon_{00}$ and $\epsilon_{20}$ for the para species, and $\epsilon_{10}$
and $\epsilon_{11}$ for the ortho species. Applying the formulae above, the para and ortho specific
heats are:
\begin{equation}
C_p=N k_B \frac{(\epsilon_{20}-\epsilon_{00}) e^{-\beta(\epsilon_{20}-\epsilon_{00})}}{(1+ e^{-\beta(\epsilon_{20}-\epsilon_{00})})^2}
\end{equation}
\begin{equation}
C_o=N k_B \frac{2(\epsilon_{11}-\epsilon_{10}) e^{-\beta(\epsilon_{11}-\epsilon_{10})}}{(1+ 2e^{-\beta(\epsilon_{11}-\epsilon_{10})})^2}
\end{equation}

Thus, at low T, the para and ortho heats in the IC depend only on the excitation energies
$(\epsilon_{20}-\epsilon_{00})$ and $(\epsilon_{11}-\epsilon_{10})$, respectively.
Fig. 3 (a,b) shows a comparison between the para, ortho and non-equilibrium specific
heats in free space and in
the IC. Notice that, at low T, in the IC the main contribution to the net
specific heat comes from $C_{o}$, whereas in free space it comes from $C_{p}$. This
is  related to the finite excitation energy ($\epsilon_{11}-\epsilon_{10}$) between (j=1) ortho states in IC,
which are degenerate in free space (see fig.1). It can be shown that the peak in the
heat capacity
occurs at $T\approx \Delta /(3 k_B)$, where $\Delta$ is the first excitation energy present in a specific
problem. Therefore, in free space the para peak is at about 80 K, while in the IC the ortho
peak is at about 75 K.

In some environments, however, the ortho-para conversion may occur quickly (e.g. due to
magnetic impurities).
In this case, the net rotational specific heat is obtained from the
composite partition function ($Eqs. 11,12$):

\begin{equation}
C_{eq}=N k_B \frac{d}{dT}(T^2 \frac{d}{dT} \ln q_{rot})
\end{equation}
At low T, the equilibrium specific heat depends only on the $(\epsilon_{10}-\epsilon_{00})$
excitation energy:
\begin{equation}
C_{eq}=N k_B \frac{(\epsilon_{10}-\epsilon_{00}) e^{-\beta(\epsilon_{10}-\epsilon_{00})}}{(1+ e^{-\beta(\epsilon_{10}-\epsilon_{00})})^2}
\end{equation}
The equilibrium specific heats of $H_2$ and $D_2$ in the IC are shown in Fig.4. In the case of
$H_2$, there is a substantial ortho-para 
conversion peak at about 20 K, followed by a gentle rise above 75 K, corresponding to
ortho-ortho excitation. However, in the case of $D_2$, the ortho-para conversion appears
only as a small bump at about 25 K, since the molar weights of para and ortho species
are now different. At
high T, the rotational specific heat per molecule
goes to $k_B$,
the classical result for a free rotor.
 
\section{Isotope slectivity}

Let us now consider a mixture of $H_2$ and $D_2$. The equilibrium condition between
the 1D gas adsorbed in the ICs or grooves and the coexisting vapor outside the nanotube
bundles is given again by $Eq.2$: $\mu_i^{IC}=\mu_i^{vapor}$, where now $i$ stands for
$H_2$ or $D_2$. The partition functions and the chemical potentials of each species 
consist of the same factors, but their meanings are different.  For example,
the rotational partition function is given now by the composite
partition function ($Eqs.11,12$) of each isotope. Since the $H_2$
and $D_2$ species have different masses, their zero point energies will be
different, and these will give a nontrivial contribution to the selectivity (as will
differences in the rotational energies). The transverse
CM oscillations of the molecule in the IC or grooveare treated as excitations of a 2D harmonic oscillator. 
The corresponding partition function is:
\begin{equation}
q_{vib}= [\sum^\infty_{n=0} e^{-\beta \hbar \omega (n+1/2)}]^2=\frac{e^{-\beta \hbar \omega}}{(1-e^{-\beta \hbar \omega})^2}
\end{equation}
where $\omega$ is the frequency of CM oscillations, found from the force constant. The
rotational hindrance is expressed in terms of the rotational energy shifts, $\epsilon_{mj}$
and the additive constants (C) in the expansion of the rotational part of the interaction
potential \cite{kos2}. This additive (reference) energy is also diferent for $D_2$ and
$H_2$. The rotational hindrance leads to a shift of the binding energy of each species:
$E_b=-(C+\hbar \omega+\epsilon_{00})$. 

The equilibrium condition for each species ($i=D_2,H_2$) yields the ratio
of IC $(\rho_i)$ and vapor ($n_i$) densities for each species, which is then used to find the isotope selectivity.
\begin{equation}
\frac{\rho_i}{n_i}= \lambda_i^2 \frac{\exp{(-\beta \hbar \omega_i)}}{(1-\exp{(-\beta \hbar \omega_i)})^2} \frac{\exp{(-\beta C)}q_{rot}^{IC}}{q_{rot}^{free}}
\end{equation}
The isotope selectivity is:
\begin{equation}
S_0= \frac{\rho_{D_2}/\rho_{H_2}}{n_{D_2}/n_{H_2}}
\end{equation}
$Eq.20$ implies that at low T the main contribution
to selectivity comes from the difference between the binding energies
\begin{equation}
S_0 \approx \frac{1}{2} \exp{[-\beta(C_{D_2}+\hbar \omega_{D_2}+\epsilon_{00}^{D_2}- C_{H_2}-\hbar \omega_{H_2}-\epsilon_{00}^{H_2})]}=\frac{1}{2} exp{[\beta(E_b^{D_2}-E_b^{H_2})]}
\end{equation}

Fig.5 displays the results of our calculations for isotopic selectivity in the ICs and grooves
of an (18,0) nanotube bundle,
and on graphite (for comparison). All three surfaces favor $D_2$.
However, the isotope selectivity in the IC and groove is much larger than that
on graphite and within the tubes \cite{jo5}. Even at 300 K the selectivity in the
IC and groove is substantial, due to the large difference between the $D_2$ and $H_2$
binding energies. This large difference comes primarily from the zero point energy (but the difference between the rotational energies also contributes).
Table 1 presents our results for the different contributions to the  the binding energies.
The difference between $D_2$ and $H_2$ binding energies is $\sim$ 400 K in the IC and
$\sim$ 170 K in the groove, much larger than the  difference of $\sim$ 35 K on graphite \cite{vid}. When exponentiated, these differences
lead to the very large
selectivities found in the IC and groove. We may compare our results with previous  calculations.
Challa {\it et al.} have studied isotope selectivity at variable pressure in the interstices 
of (10,10) nanotube bundle, using Ideal Adsorbed Solution Theory(IAST) and Path Integral 
Grand Canonical Monte Carlo simulations (PI-GCMC) \cite{jo5}. At low pressures,
the IAST
and PI-GCMC results converge to the analytic zero pressure results.
In their calculation, only the transverse zero-point
energy  is considered, i.e. rotational motion is ignored. The selectivity calculated by them at 20 K and very low
pressure ($10^{-13}$ atm) is about 700.
(10,10) and (18,0)
nanotubes have similar diameters (13.6 and 13.8, respectively), so the interstices of 
a closed packed nanotube bundle are similar, too ($\sim$ 6 $\mathring{A}$ diameter). Recently, Hathorn {\it et al.} \cite{hath} have estimated the quantum effects due to the restriction of the rotational motion of $H_2$ and $D_2$ adsorbed within a single nanotube. Their results reveal significantly enhanced rotational separation factor(selectivity), which at $20$ K is of order $10^3$. 
The zero pressure isotope selectivity calculated by us at 20 K, in the interstices
of bundles of (18,0) carbon nanotubes is of order $10^9$.
Our results differ to such a large degree since we considered both zero point
and rotational contributions to selectivity and  we used 
a slightly deeper, solid-fluid potential. The stronger the confinement,
the larger the
difference between the isotope zero point energies and between the rotational energies.
Previous calculations
inside (5,5) nanotubes at 20 K yielded much lower isotope selectivities \cite{gor}. At low pressure,
the computed selectivity of tritium to $H_2$ is 23, and that of tritium to $D_2$ is 1.7;
hence $S_0(D_2/H_2)=13$. Thus, even if the inner channels of (5,5) nanotubes have
about the same radius (3.3 $\mathring{A}$) as the ICs of (18,0) nanotube bundles (3 $\mathring{A}$), the
computed selectivity in the IC is much larger due to the different arrangement of the carbon atoms
in the IC, which yields greater confinement.
 
\section{Isosteric heat}

The isosteric heat is defined as:

\begin{equation}
Q_{st}= -[\frac{d(ln{P})}{d \beta}]_N
\end{equation}
where P is the pressure of the vapor outside the nanotube bundles. In our model, at low
density, the dependence of P on T can be found from
the equilibrium condition
$(Eq. 2)$ using the ideal gas formula: $n=\beta P$. Finally, the isosteric heat
is:
\begin{equation}
Q_{st}=2 k_B T-\hbar \omega - C + (\frac{d}{d \beta})[ln \frac{(q_{rot}^{IC}/q_{rot}^{free})}{(1-e^{-\beta \hbar \omega})^2}]  
\end{equation}

At low T, the excitations can be neglected and 
the term involving the derivative with respect to $\beta$
goes to $-\epsilon_{00}-(\epsilon_{10}-\epsilon_{00}) e^{-\beta(\epsilon_{10}-\epsilon_{00})}+2B e^{-2\beta B}$, where $\epsilon_{00}$
and $\epsilon_{10}$ are the first two rotational energy levels.
Therefore, at low T, the isosteric heat is a
measure of the binding energy:
\begin{equation}
Q_{st}=E_b+2 k_B T
\end{equation}
since $E_b=-(C+\hbar \omega+\epsilon_{00})$.

The isosteric heat can be found experimentally
by taking the difference between two nearby isotherms. Recently, there have been
reported experiments of adsorption of $D_2$ and $H_2$ in carbon nanotube bundles [33], which
show isosteric heats in nanotube bundles to be a factor of 1.5 (for $H_2$) to 1.8 (for $D_2$)
larger than those on graphite.
This means that the
$D_2$ and $H_2$ binding energies are almost twice as large as those on graphite. Moreover, the
difference between isotope isosteric heats, which again is related to the difference between isotope
binding energies, was found to be about 200 K, much greater than the difference of
about 35 K on graphite \cite{vid}. This is a mark of the greater confinement of molecules
adsorbed in nanotube bundles than on the surface of  graphite; the  confinement leads to the enhanced
separability of isotopes found in the previous section.

We performed calculations of the isotope isosteric heats in the low density limit ($Eq .24$).
Table 2 and Figs. 6(a),(b) compare the results of the calculated isosteric heats 
and the low coverage experimental data at T=85 K.
The experimental procedure of Wilson {\it et al} does not reveal where
in the nanotube bundle the adsorption occurs. Since the experimental 
isosteric heats are much larger than on graphite, one assumes that the adsorption environment
should be either the IC or the groove site. Our calculations of the isosteric heats
in both IC and groove are lower than the experimental results.
The calculated difference between $D_2$ and $H_2$ isosteric
heats (374 K) in the IC is much larger than the one found experimentally, but the
difference (137 K) in the groove  is closer to the experimental
difference (200 K). In addition, the calculated binding energies(see Table I) for $D_2$ and $H_2$ in the grooves are consistent with the experimental results and show that the adsorption on the grooves is much more likely than on the IC at low T.  Assuming the accuracy of our approximations, we believe
that our results are compatible with an adsorption in the groove, rather than the IC. This
is also experimentally plausible, since the external grooves are more accessible for
adsorption than the ICs. Indeed, the ICs may be completely blocked, preventing any adsorption there.

\section{Summary and conclusions}

We have investigated the spin (ortho versus para) and the isotope ($D_2$ versus $H_2$)
selectivity at low coverage in the IC and groove channel of (18,0) nanotube bundles.
The rotational hindrance of 
molecules in the IC and groove site induces shifts and splittings in the rotational
spectrum, enabling the nanotube bundles to preferentially adsorb $H_2$ ortho molecules
and $D_2$ para molecules. Our calculations show substantial
spin selectivity
for temperatures up to $\sim$ 100 K in the IC and up to $\sim$ 75 K in the groove.
At low temperatures, the alterations of the rotational spectrum in the IC induces
new features in 
the ortho heat capacity, as the ortho-ortho peak. The $H_2$ equilibrium heat capacity
exhibits a distinctive ortho-para conversion peak at $\sim$ 20 K. The different features
of non-equilibrium and equilibrium heat capacities may be a way to check the existence of
the ortho-para conversion of $H_2$ and $D_2$ adsorbed in nanotube bundles.
In the case of the isotope selectivity in nanopores,
zero-point motion favors the heavier isotope \cite{jo5}, and the rotational motion
enhances the preferential adsorption of the heavier isotope \cite{kos2,hath}. 
We find substantial isotope
selectivity even at temperatures
as high as 300 K in the IC and groove. At 20 K, our calculation at low coverage
(consequently zero pressure) yield isotope selectivities of order $10^9$ in the
IC, orders of magnitude larger than the low pressure limit found in the calculations of Wang {\it et al.} and Challa {\it et al.}.
Our much larger isotope selectivity is a consequence
of the stronger confinement in the IC, which yields larger zero point and rotational
energies. The difference between the $D_2$ and $H_2$ isosteric heats in the groove 
($\sim$ 135 K) is close to the value found
experimentally ($\sim$ 200 K). We conclude that carbon nanotube bundles are ideal surfaces
for spin
and isotope separation, providing a potential technology of quantum sieving. 

We acknowledge stimulating discussions with J. K. Johnson, M. M. Calbi, T. Wilson,
M. J. Bojan, O. Vilches, P. Sokol, D.
Narehood and D. Stojkovic. This research has been supported by 
Petroleum Research Foundation of the American Chemical Society and Air Products and Chemicals, Inc.

\newpage
%\begin{references}

\newpage

\begin{center}
\begin{table}[h]
\caption{The zero point energies $(\hbar \omega)$, rotational energy references (C)
and shifts $(\epsilon_{00})$ and binding energies ($E_b$) (expressed in Kelvin). $D_2-H_2$
means the difference between the corresponding values. }
\setlength{\tabcolsep}{0.2cm}
\renewcommand{\arraystretch}{1.5}
\begin{tabular*}{14.9 cm}[t]{|c|c|c|c|c|c|c|c|c|c|}\hline

   &    $\hbar \omega^{IC}$& $\hbar \omega^{groove}$& $C^{IC}$& $C^{groove}$& $\epsilon_{00}^{IC}$& $\epsilon_{00}^{groove}$& $E_b^{IC}$& $E_b^{groove}$& $E_b^{graphite}$ \\ \hline

$D_2$&   762& 266& -1445& -970& -133& -62& 816& 766& 517 \\ \hline

$H_2$&   1077& 370& -1278& -880& -206& -93& 407& 603& 482 \\ \hline

$D_2-H_2$&	-315& -114& -166& -90& 73& 31& 409& 163& 35 \\ \hline

\end{tabular*}
\end{table}
\end{center}

\begin{center}
\begin{table}[h]
\caption{The isotope
isosteric heats (Q) in the IC, groove and experimental results (expressed in Kelvin) }
\setlength{\tabcolsep}{0.4cm}
\renewcommand{\arraystretch}{1.5}
\begin{tabular*}{8.2 cm}[t]{|c|c|c|c|}\hline

   &     $Q_{calc}^{IC}$&  $Q_{calc}^{groove}$&  $Q_{exp}$ \\ \hline

$D_2$&   991& 901& 1100$\pm$90 \\ \hline

$H_2$&    617& 764&  900$\pm$70 \\ \hline

$D_2-H_2$&	374& 137&  200$\pm$110 \\ \hline

\end{tabular*}
\end{table}
\end{center}

\newpage

{\LARGE \bf Figure captions}

Fig. $1$. $H_2$ rotational spectrum in free space and in IC. The number of lines in a level
represents the degeneracy of that energy level. o and p mean ortho and para, respectively.

Fig. $2$. $D_2$ and $H_2$ spin selectivity in IC and groove channel as a function of T

Fig. $3$. Ortho (dashed line), para (dotted line) and non-equilibrium (full line)
specific heat (without spin equilibration) of $H_2$
molecules (a) in IC, (b) in free space

Fig. $4$.  The equilibrium rotational specific heat of $H_2$ (full line)
and of $D_2$ (dashed line)

Fig. $5$. Isotope selectivity in the IC (full line), groove (dashed line) and on graphite (dotted
line).

Fig. $6$. The calculated $H_2$ (dotted line), $D_2$ (dashed line) and the difference $D_2-H_2$ (full line) isosteric
heats (a) in the IC and (b) in the groove. The experimental values at 85 K
are shown in symbols:  circle for $D_2$, square for $H_2$ and triangle for the difference $D_2-H_2$.


\begin{thebibliography}{99}
\bibitem{energy}
The reference energy is the first term in the expansion
of the rotational part of the confining potential:
$V=C+C_2 cos^2 \theta+C_4 cos^4 \theta$. The minimum of the potential ($C+C_2+C_4$)
is the same for 
$H_2$ and $D_2$, but their expansion terms $C_i$ are different since they have been derived after averaging the holding potential over the  zero point motion of $H_2$ and $D_2$. For more details see Ref. 35.
\end{thebibliography}
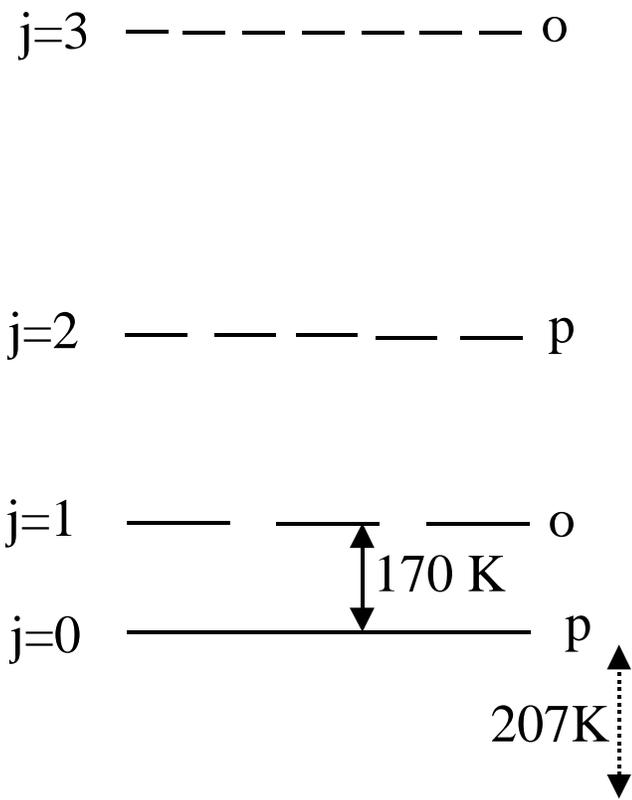
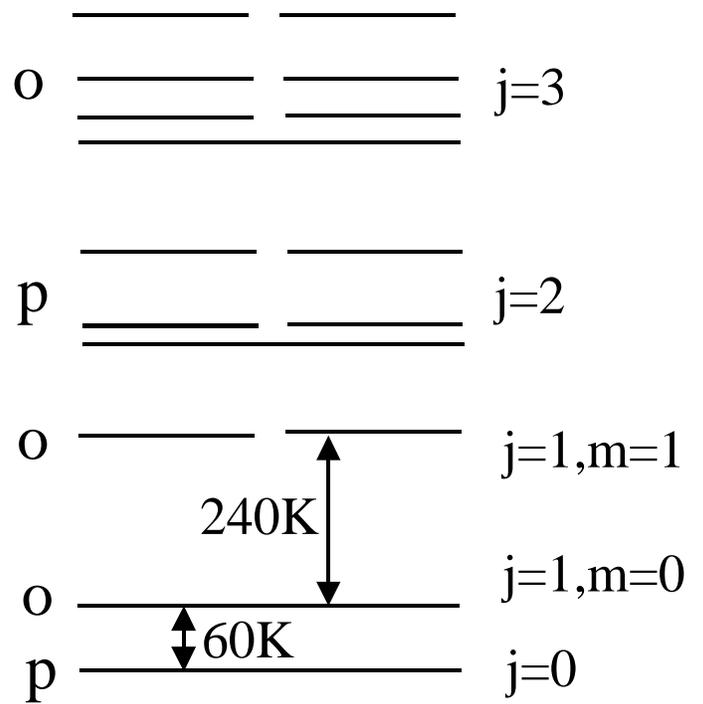
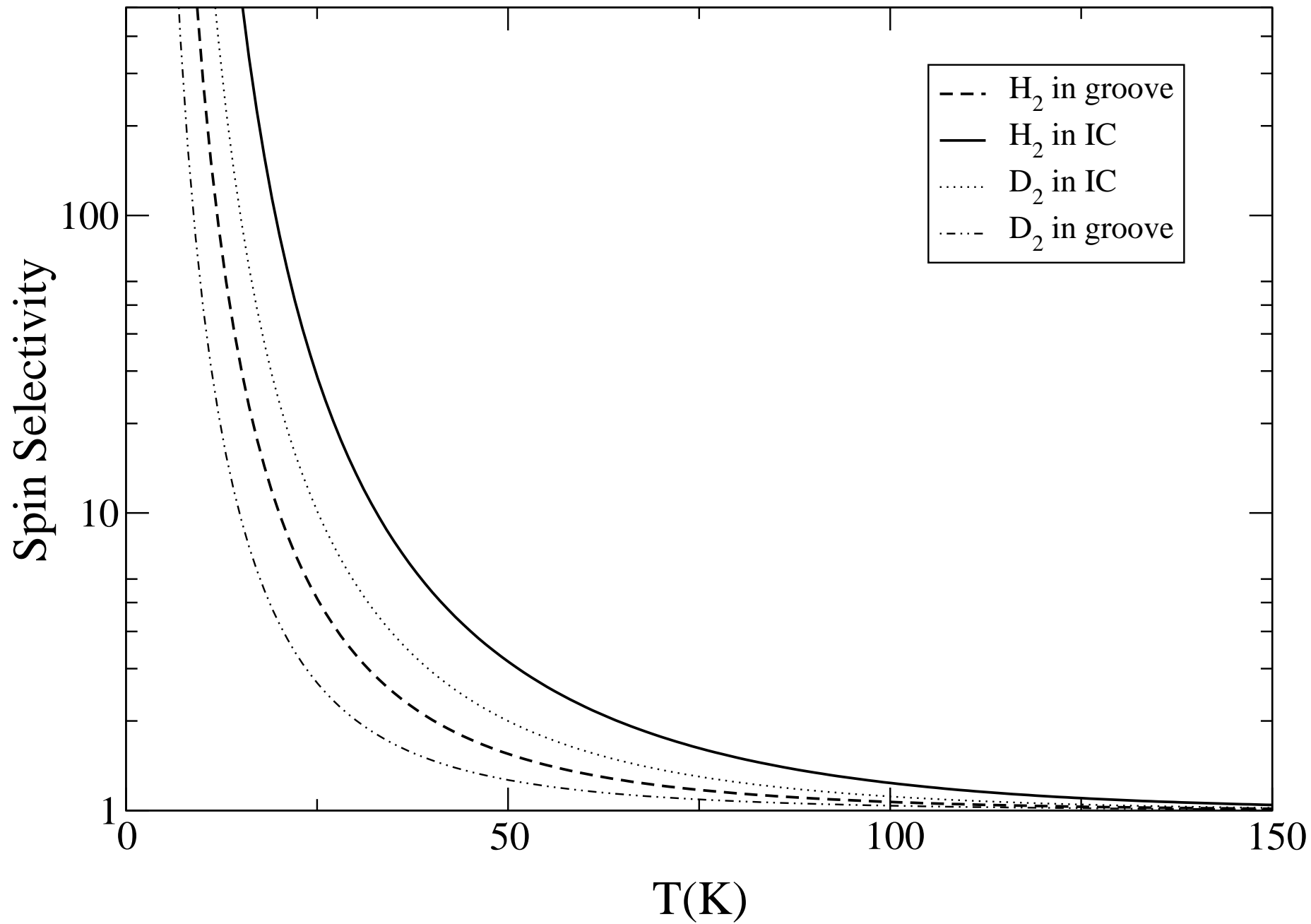
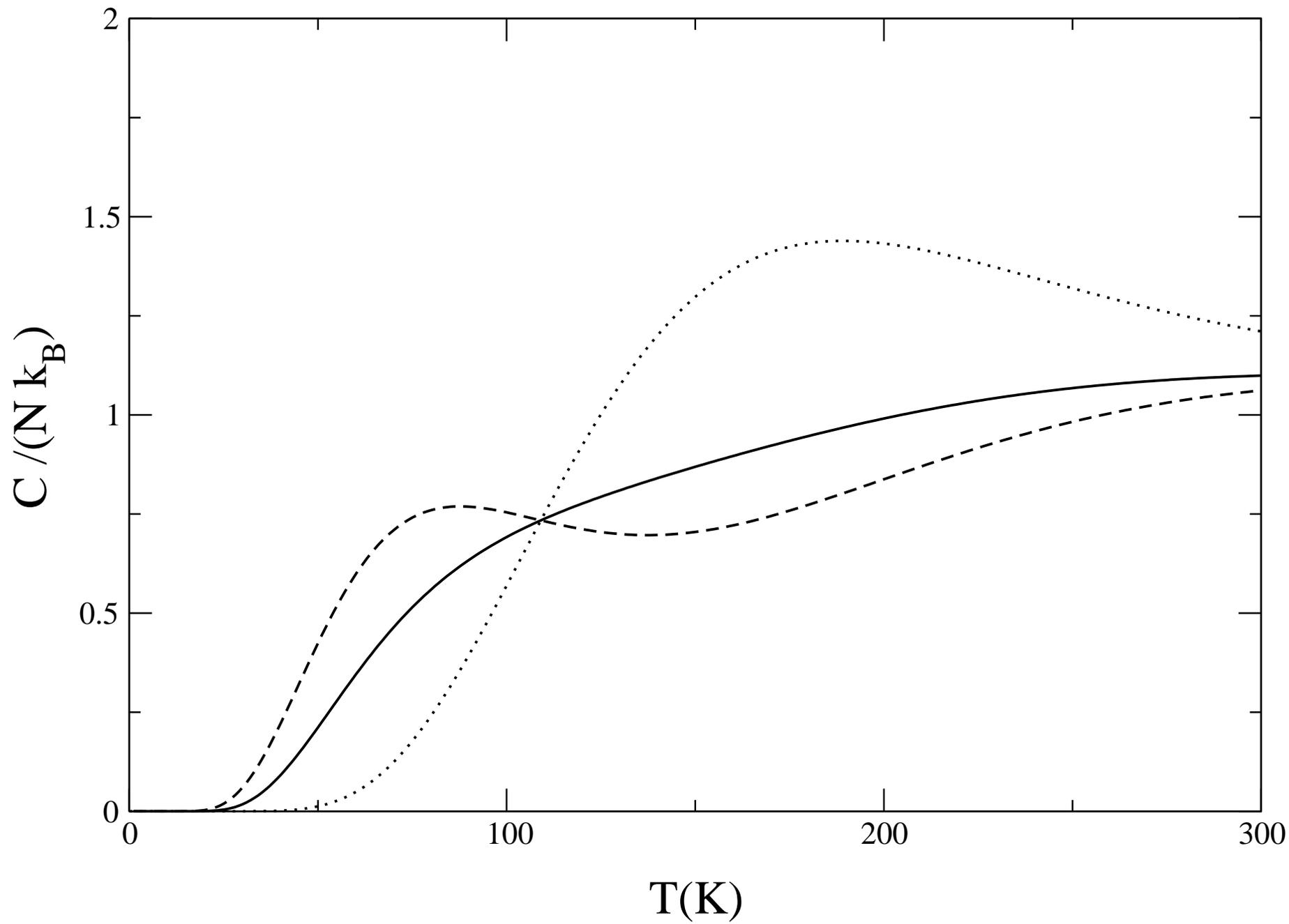
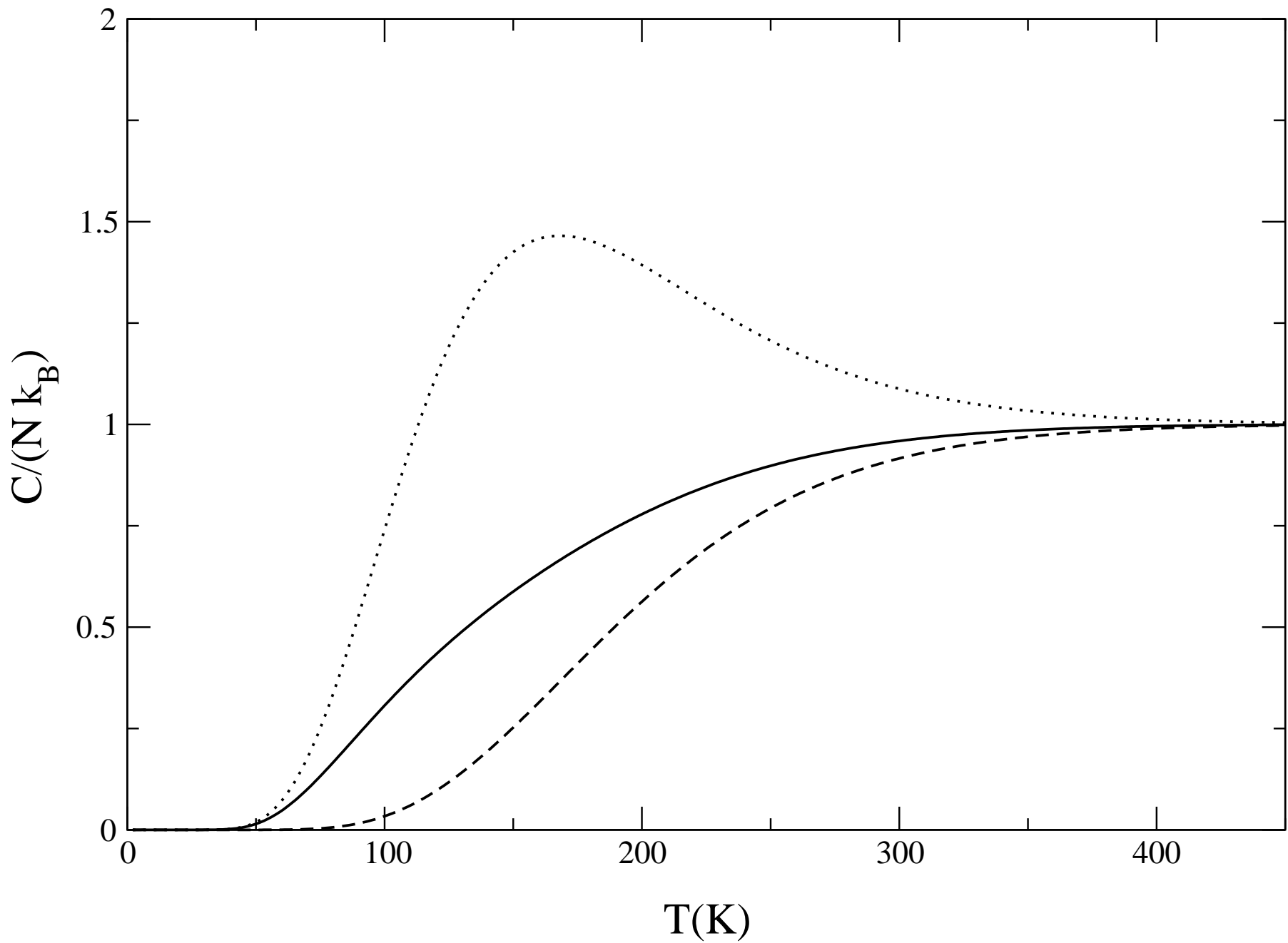
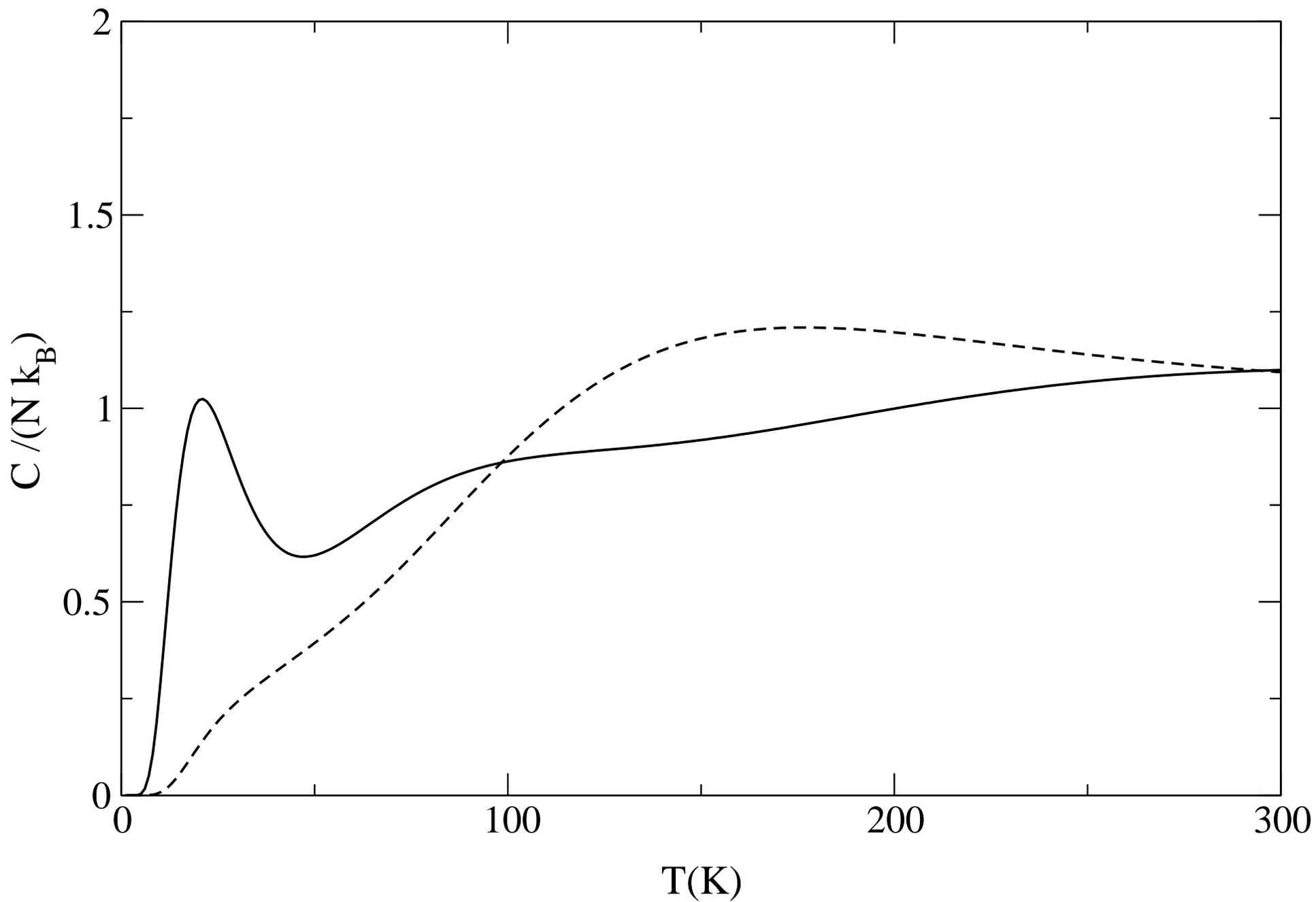
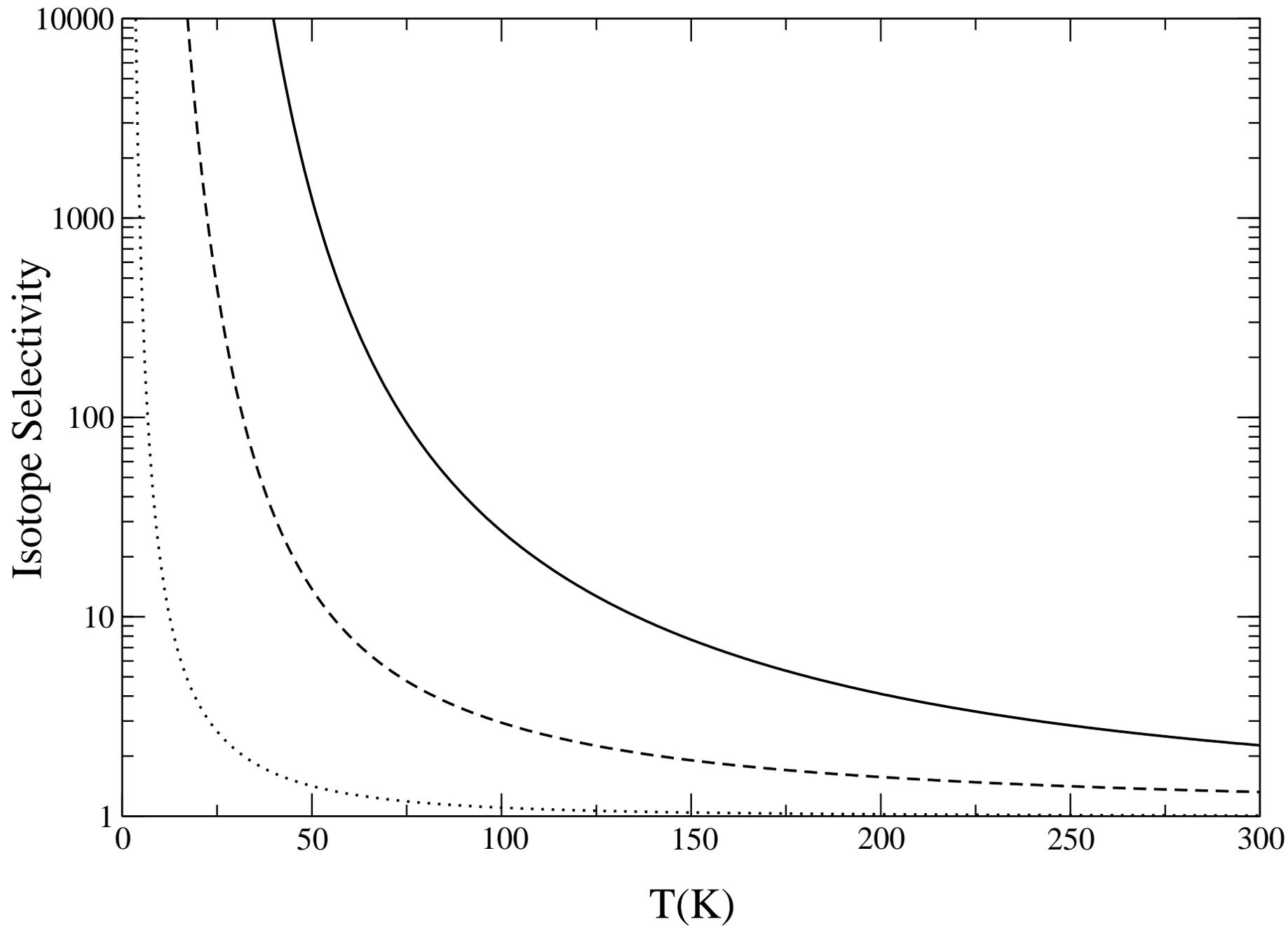
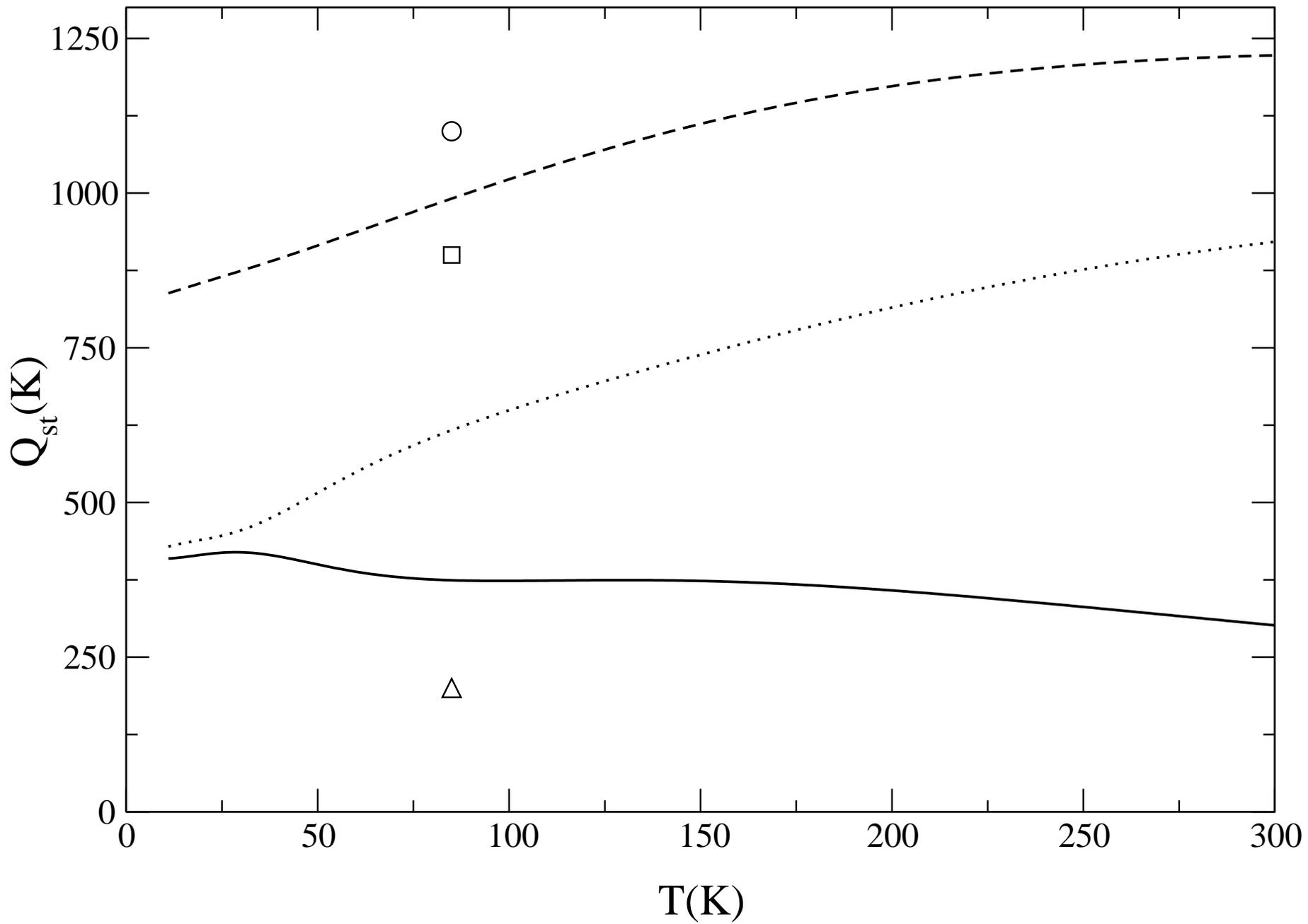
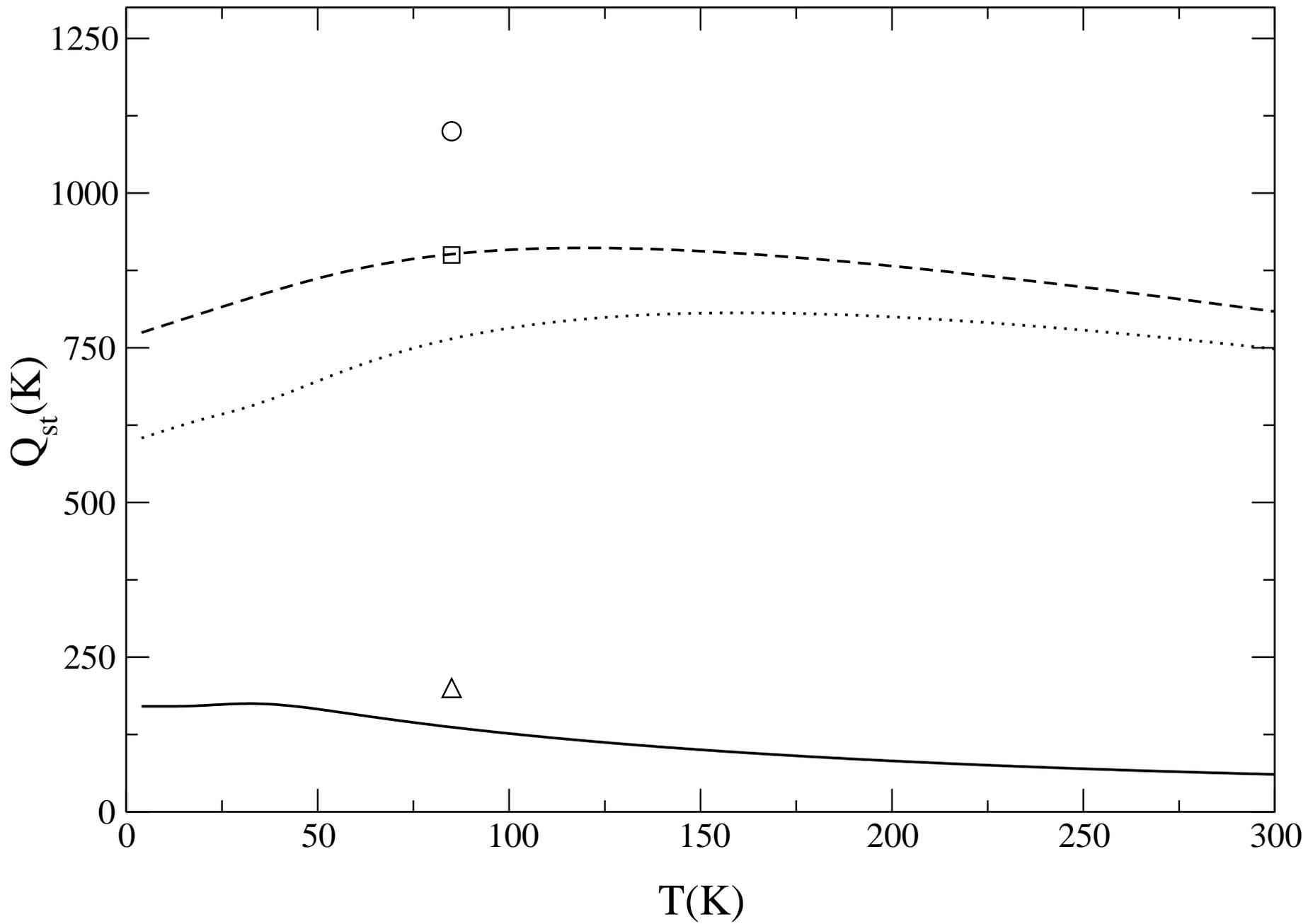
\end{document}